# Superconducting Transport in single and parallel double InAs Quantum Dot Josephson Junctions with Nb-based Superconducting Electrodes

Shoji Baba[1,a)], Juergen Sailer[1], Russell S. Deacon[2,3], Akira Oiwa[4], Kenji Shibata[5,6], Kazuhiko Hirakawa[5,7] and Seigo Tarucha[1,2,8,9]

[1] *Department of Applied Physics, University of Tokyo, 7-3-1 Hongo, Bunkyo-ku, Tokyo 113-8656, Japan.*

[2] *Center for Emergent Matter Science (CEMS), RIKEN, Wako, Saitama 351-0198, Japan.*

[3] *RIKEN Advanced Science Laboratory, 2-1 Hirosawa, Wako, Saitama 351-0198, Japan.*

[4] *The Institute of Scientific and Industrial Research, Osaka University, 8-1 Mihogaoka, Ibaraki, Osaka 567-0047, Japan.*

[5] *Institute of Industrial Science, University of Tokyo, 4-6-1 Komaba, Meguro-ku, Tokyo 153-8505 Japan.*

[6] *Department of Electronics and Intelligent Systems, Tohoku Institute of Technology, Sendai 982-8577, Japan.*

[7] *JST CREST, 4-1-8 Hon-cho, Kawaguchi-shi, Saitama 332-0012, Japan.*

[8] *INQIE, The University of Tokyo, 4-6-1 Komaba, Meguro-ku, Tokyo 153-8505, Japan.*

[9] *QPEC, The University of Tokyo, 7-3-1 Hongo, Bunkyo-ku, 113-8656, Japan.*

## ABSTRACT

We report conductance and supercurrent measurements for InAs single and parallel double quantum dot Josephson junctions contacted with Nb or NbTiN superconducting electrodes. Large superconducting gap energy, high critical field and large switching current are observed, all reflecting the features of Nb-based electrodes. For the parallel double dots we observe an enhanced supercurrent when both dots are on resonance, which may reflect split Cooper pair tunneling.

___________________

a) Electronic mail: baba@meso.t.u-tokyo.ac.jp

**TEXT**

Inherently low-dimensional semiconductors, such as nanowires, nanotubes, and self-assembled quantum dots, are attractive research materials, because their natural geometries are suited to study anisotropic quantum effects, and they are easily combined with superconductors or ferromagnets to investigate hybrid quantum effects. In particular, by directly contacting self-assembled quantum dots (SAQDs) with normal metal electrodes, gate-tunable electronic properties of confined electrons have been revealed, such as anisotropic and tunable spin-orbit interaction[1,2] and g-factor[2,3], and tunable Kondo-effect[4,5]. For hybrid devices contacted with superconducting electrodes various kinds of proximity induced quantum transport have been revealed, including non-dissipative supercurrent[5], 0-π junction phase transitions[6], spectroscopy of Andreev energy levels[7,8], and competition between superconducting proximity effect and Kondo screening[9,10]. In most of these studies aluminum (Al) is used for the superconducting contacts primarily because it is an easy material for making small but robust superconducting junctions with well-established nano-fabrication techniques. Though experiments on QD Josephson junctions (QDJJs) with Al electrodes have been very successful to date, the parameter space is restricted to low magnetic field and low temperature and in most devices the superconducting energy gap maybe small compared with other energy scales. A superconductor with a large superconducting gap and a high critical field would extend the parameter space that can be studied with QDJJs and open new possibilities to study devices in high magnetic fields. One area in which high $T_c$ leads may be of benefit is the study of Cooper-pair beam splitting with two quantum dots parallel coupled to a superconductor, which provides a source of non-local entangled electron pairs for quantum information processing and quantum electron optics. Although Cooper-pair splitting is confirmed in several studies[11–14] with InAs nanowire or carbon nanotube devices, to prove the maintenance of the spin-singlet correlation after pair-splitting remains challenging. We have recently observed increased switching current for parallel double quantum dot Josephson junctions (DQDJJs) with Al electrodes when both dots are on resonance, indicating split but correlated Cooper pair tunneling through the two dots[15]. To further study the spin correlation in the split tunneling of paired electrons it is proposed to detect the phase factor in two path interference[16–18], for which parallel double QDJJs made of superconducting electrodes with a high critical field are necessary to control the relative phase between the two paths. Fabrication of InAs SAQD junctions with contacts from Nb and its alloys instead of Al will enable such experiments. However, Nb can be a difficult material with which to fabricate nanoscale superconducting structures. Being a refractory metal with high melting point it is unsuitable for electron beam deposition without significant efforts to reduce the risk of oxygen contamination (for example due to outgassing resist) which can drastically effect the superconducting properties for small structures. Indeed there have been no reports on fabrication of QDJJs with InAs SAQDs contacted by electrodes made from Nb or other large superconducting gap materials to date, whereas there are a number of reports on InAs or InSb nanowires coupled to such superconductors [19–26]. In this study we report the fabrication and characterization of Nb-based self-assembled QDJJs, in which gate-tunable InAs single and double SAQDs are utilized. For the single QDJJs (SQDJJs) we produce transparent contacts using sputtered NbTiN and clearly observe a flow of supercurrent through the dots. The measured superconducting gap energy is much higher than that of conventional SQDJJs with Al electrodes. Consistently, a high critical field and a large critical current are observed. We also fabricated DQDJJs with two quantum dots in parallel and closely spaced but neither quantum mechanically nor electrostatically coupled with each other between two NbTiN electrodes. We observe a signature of increased supercurrent, which may be due to split tunneling of Cooper pair electrons through the two dots.

SQDJJs and DQDJJs were fabricated in the following way. InAs SAQDs are grown on the surface of a GaAs substrate by molecular beam epitaxy. The area density of the dots is $1.5\text{-}2 \times 10^9/cm^2$, and the typical dot diameter ranges from 50 to 200 nm. During growth, a Si-doped layer is included 300nm below the surface to be used as a global backgate. Since the InAs SAQDs are randomly distributed on the GaAs surface, precise alignment is required for placing the contacting electrodes in the right positions. For this purpose we preselect QDs using images taken with a low-damage SEM (JSM-7500FA) and obtain their coordinates in relations to previously fabricated nanoscale markers. We note that devices made using a conventional SEM were unable to achieve working devices despite accurate alignment and so the use of a gentle beam or low damage SEM system (or AFM observation in other studies) was critical to limit damage to the device. The typical misalignment of the deposited electrodes was 10 to 20 nm. We select nominal 150nm-diameter quantum dots to make SQDJJs (See Fig.1(a)). To fabricate our contacts two different lead materials have been used. In one case we deposit Ti/Nb/Al contacts (3/100/10nm) using electron beam deposition for Ti/Al and DC magnetron sputter deposition for Nb. In these devices Ti was found to be necessary to achieve transparent superconducting contacts and the Al layer was included as a barrier to oxidation of the Nb layer. Alternatively we deposited NbTiN (110 nm) using reactive DC magnetron sputtering. In both approaches source and drain electrodes with a gap of ~50nm were patterned over the QD using PMMA resist and conventional liftoff. In order to improve the metal contact quality an oxidized layer which covers the dot surface is removed using in-situ Ar plasma etching before depositing the contact metal(s). Close to the junction local sidegates are placed to control the electrostatic potential of the dot. Conductivity of the SQDJJs and DQDJJs was measured by applying either current or voltage bias across the junction, while gate-tuning the QD's electrochemical potential. At room temperature ~90% of the fabricated SQDJJs had a reasonable resistance<300k$\Omega$, whereas for DQDJJs the yield was ~50%, due to smaller overlap between leads and each QD.

We first focus on the characterization of the SQDJJs with Ti/Nb contacts. We measured the differential conductance of SQDJJs under voltage bias using a conventional lock-in technique with frequency of 37 Hz and ac excitation voltage of $V_{ac}$=5 μV. The samples were immersed in a $^3$He cryostat with base temperature of ~0.3K. Fig. 1(b) shows the differential conductance as a function of source-drain voltage $V_{sd}$ and backgate voltage $V_{bg}$ measured for an out-of-plane magnetic field $B_z$ = 0T. At around $V_{sd}$=±1.1mV, conductance peaks due to direct quasi-particle tunneling are observed. Coulomb diamonds are not clearly visible due to the presence of these peaks, though from the data taken in high magnetic field of ~4T we obtained the charging energy $U$ = 2-3 meV, and the level spacing $\xi$~1 meV[27], which are comparable with those of previous measurements for InAs SAQDs of the same size. Typical critical temperature and critical field of the contact leads are evaluated as ~7.0K and ~4.5T. The critical temperature measured is smaller than that measured on separately fabricated thin films which show $T_c$~8.5K.

Fig. 1(c) shows a d$I$/d$V$ trace of Fig. 1(b) at $V_{bg}$=1.2 V and its out-of-plane magnetic field ($B_z$) dependence. At $B_z$=0 T conductance peaks due to direct quasi-particle tunneling are observed. The separation of these peaks decreases as $B_z$ increases, and they disappear near $B_z$=5T. From these peaks at $B_z$=0 T we derive the gap of $\Delta$~0.55 meV using the relation 2$\Delta$=e$V_{sd}$. The $\Delta$ obtained here is about one-third of the value for bulk Nb (~1.5meV). The observation of a reduced gap and $T_c$ may be due to degradation of the sputtered material possibly caused by oxidization, which is expected to be more severe in the periphery of the electrodes, including the QD-lead interfaces.

Although we confirmed contact of Nb electrodes to InAs SAQDs, we could not detect supercurrent flow in any of the devices we fabricate. The combination of electron beam deposited Ti contact/adhesion layer and sputtered Nb maybe non-ideal as the sputtered layer can easily cover the edge of the Ti layer. As a result the leading edge of the contact to the dot maybe Nb rather than Ti/Nb affecting the junction transparency. To improve this issue we studied sputtered NbTiN contacts which did not require the deposition of additional contacting films. From here, all experiments were performed in a He$^3$-He$^4$ dilution refrigerator with base temperature of 50 mK and fridge lines with homemade copper powder, 2$^{nd}$ order RC and pi filters. The measured differential conductance d$I$/d$V_{sd}$ at $B_z$ =0 T is shown in Fig. 2(a). The conductance is generally high and Coulomb oscillation is unclear, both suggesting the strong coupling between the QD and contact electrodes. In addition a clearly pronounced zero-bias peak(ZBP) is observed in the entire gate region and attributed to the supercurrent.

The measured differential resistance d$V_{sd}$/d$I_{sd}$ at $B_z$ = 0 T is shown in Fig. 2(b). Blue color in the center shows a low resistance indicating the supercurrent flow. The typical $V$-$I$ trace is shown in Fig. 2(c). We attribute finite resistance around zero current to thermal phase diffusion[28]. Switching current $I_{sw}$ as defined from the threshold behavior is modulated ranging from 2.5nA to 6nA as a function of gate voltage $V_{bg}$. This $I_{sw}$ is much higher than that (~1 nA) for QDJJs with Al electrodes we previously measured[9]. In Fig. 2(a), integration of $G$≡d$I_{sd}$/d$V_{sd}$ with respect to $V_{sd}$ across the ZBP ($V_{ZBP-}$ ~ $V_{ZBP+}$ : -0.1mV~0.1mV ) gives a current $I_{int}$ as

$$I_{int} \equiv \int_{V_{ZBP-}}^{V_{ZBP+}} G \, dV_{sd}$$

,which is shown in Fig. 2(d) and has close agreement with the directly measured $I_{sw}$.

Shown in Fig. 3 is the magnetic field dependence of differential resistance d$V$/d$I$ measured using current bias with $V_{bg}$=0.5 V for in-plane magnetic field $B_x$ in (a) and for the out-of-plane magnetic field $B_z$ in (b). Supercurrent is detected in the presence of an in-plane magnetic field as high as $B_x$ =1 T, which is the upper limit of the superconducting vector magnet used in this measurement. On the other hand, for the out-of-plane magnetic field supercurrent is quenched already at $B_z$ =0.2 T. The anisotropy in the critical magnetic field arises from the anisotropic geometry of the superconducting thin film used for contacts which has a thickness of only ~110nm such that critical field is only weakly influenced by the in-plane magnetic field $B_x$.

Finally we will discuss DQDJJs with NbTiN electrodes, which were prepared in the same way as the SQDJJs but with two closely spaced InAs dots (see Fig. 4(a)). The inset of Fig. 4(a) shows a low damage SEM image of the double dot before the metal deposition for the contacts. The two dots with diameter of ~100nm are separated from each other by the center-to-center distance of ~100 nm. Two sidegates, sgr and sgl, are placed to control each dot separately.

Fig. 4(b) shows the zero-bias differential conductance, dI/dV vs. $V_{sgr}$ and $V_{sgl}$. Two types of conductance peaks with different inclinations (highlighted by dashed lines) indicate Coulomb resonances of the two QDs. Note that the resonance peaks (white and green dashed lines) of QD1 are much weaker than those for QD2, indicating a weaker tunnel coupling to the leads. Charging energies are comparable for the two dots with $U_1 \sim U_2 =$ 1 to 2 meV, and $\Delta \sim 0.95$ meV estimated from measurement of Coulomb diamonds[27]. We also performed *V-I* measurements in the gate voltage region of Fig. 4(b). Examples of typical data are shown in Fig. 4(d) taken at five different gate voltage positions marked by stars along the green dashed line in Fig. 4(b). Relative positions of these are shown in Fig. 4(c) as a schematic illustration: both dots on resonance in red (D) and orange (B) and QD1 only on resonance in cyan (A), blue (C), and black (E), respectively. Here the resistance of 7.5 kΩ (extracted from an *V-I* trace with the minimum zero-bias resistance), which is attributed to imperfect transparency and thermal phase diffusion[28], is subtracted from all data for clarity. The *V-I* traces show weak but discernable features of enhanced conductance near the origin and current thresholds where the resistance starts to increase when QD2 is on resonance (see the red and orange traces), suggesting the existence of an incompletely formed supercurrent, while no such feature is observed when only QD1 is on resonance, indicating suppressed or no direct local Cooper pair tunneling through QD1. The threshold like feature is completely suppressed when QD2 is off resonance, independent of whether QD1 is on resonance or not. Fig. 4(e) shows the same plots as in Fig. 4(d) but taken at the gate voltage positions marked by stars along the QD2 resonance line on the right: both dots on resonance in red (D), and only QD2 on resonance in magenta (F) and green (G). When QD2 is on resonance all traces show a threshold like feature suggesting a flow of supercurrent. The threshold current is increased when both dots are on resonance (red trace, or trace (D)). The tendency is observed for the other QD2 resonance line as shown in Fig. 4(f). Considering that there is no supercurrent detected when only QD1 is on resonance, this increased threshold or switching current may be assigned to the contribution from split tunneling of Cooper-paired electrons through the two dots as we reported previously for InAs DQDJJs with Al superconducting electrodes[15]. We note that this increase of the threshold is not from a simple resonance of QD1, because the magnitude of conductance enhancement is in the order of $10^{-5}$ to $10^{-4}$S, which is much higher than QD1 resonance peaks(~$10^{-6}$S).

In conclusion, we have fabricated InAs SAQD Josephson junctions with Nb-based electrodes. The observed superconducting transport properties resemble those from previous studies, but physical quantities to characterize the superconductivity, such as critical current and critical field, are higher than those measured in QDJJs with InAs SAQDs contacted by Al electrodes. We have also fabricated double-dot Josephson junctions. From the zero-bias conductance measurement with the two side-gate voltages we confirm the stability diagram of the parallel double dot. In the current bias measurements we observe enhanced conductance near zero-current and its suppression at a finite current, suggesting a flow of supercurrent when one of the two dots is on resonance. In addition, we observe a tendency of increased current threshold when both dots are on resonance, providing strong evidence for the presence of split Cooper pair tunneling. These results are encouraging for future experiments in high magnetic fields to probe non-local spin-entanglement of split Cooper-pairs. We would also like to note that devices comprised of arrays of QDs coupled to a superconductor have been proposed for the study of Majorana fermions[29], and the fundamental fabrication technique for such a device is established in this study.

# of words : TEXT+CAPTIONS+FIGURES+EQUATION=2443+329+705+16=3493(words)

## ACKNOWLEDGEMENTS

Part of this work was supported by Grant-in-Aid for Scientific Research(S) (No. 26220710), Grant-in-Aid Research A from Nos. 25246004 and 25246005 and Innovative areas (No. 26103004), Grants-in-Aid for Young Scientists B (No. 26790008), Grants-in-Aid from JSPS (Nos. 25600013, and 26706002) MEXT, Grant-in-Aid for Scientific Research on Innovative Areas "Science of hybrid quantum systems" (No. 2703), Project for Developing Innovation Systems of MEXT, JST Strategic International Cooperative Program (DFG-JST), FIRST program, ImPACT Program of Council for Science, Technology and Innovation (Cabinet Office, Government of Japan), MEXT Project for Developing Innovation Systems, and Japan Society for the Promotion of Science through Program for Leading Graduate Schools (MERIT). S.B. acknowledges support from Grants-in-Aid for JSPS fellows.

# FIGURES

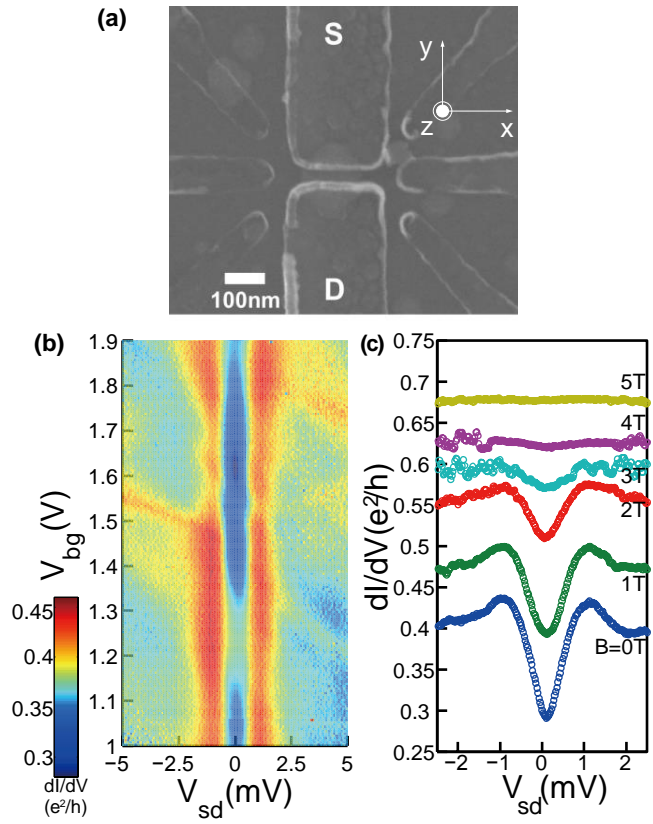

**Figure 1.** (a) A SEM image of a QDJJ device. Ti/Nb/Al source and drain electrodes are contacted to an InAs self-assembled quantum dot, whose diameter is about 150nm. Local sidegates are fabricated near the junction. (b) Plot of differential conductance d$I_{sd}$/d$V_{sd}$ as a function of backgate and source-drain voltages. The differential conductance is recorded at the base temperature of ~300mK and in the absence of external magnetic field. (c) Magnetic field dependence of conductance as a function of $V_{sd}$, measured at $V_{bg}$=1.2V. Each graph is shifted vertically by 0.07×e^2/h for clarity.

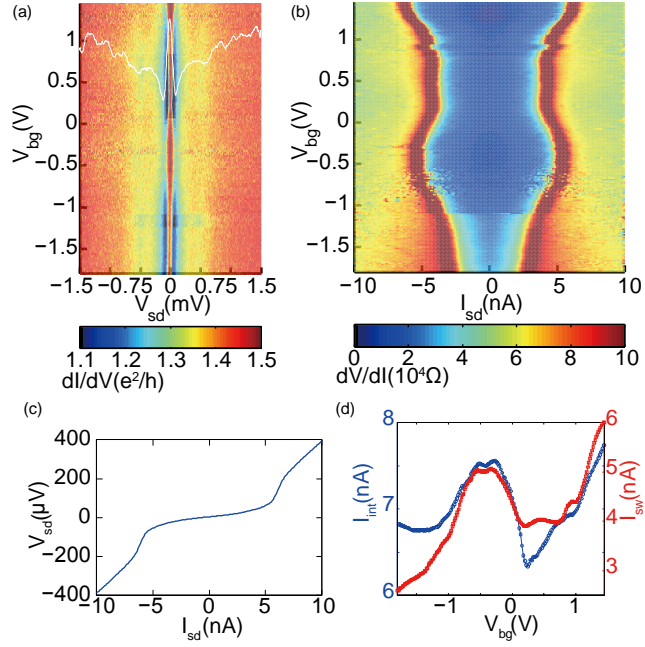

**Figure 2.** (a) Plot of differential conductance d$I_{sd}$/d$V_{sd}$ as a function of backgate voltage ($V_{bg}$) and source-drain bias voltage ($V_{sd}$). (b) Plot of differential resistance (d$V_{sd}$/d$I_{sd}$) measured under the same condition as (a). Blue part in the middle corresponds to supercurrent. Switching current is modulated in the range of 2.5~6 nA. (c) A typical I-V trace extracted from (b) at $V_{bg}$=1.2V. (d) Switching current ($I_{sw}$) and the integration of zero-bias conductance peak ($I_{int}$) for the gate region shown in (a) and (b).

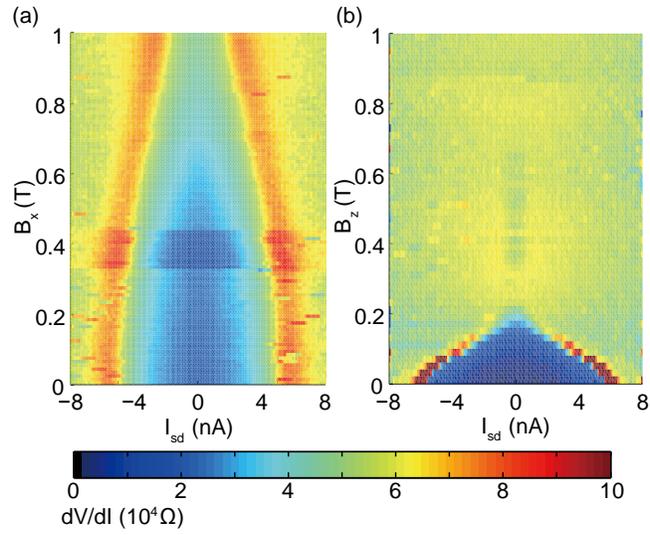

**Figure 3.** (a) Plot of differential resistance $dV_{sd}/dI_{sd}$ as a function of in-plane magnetic field $B_x$ and source-drain bias current $I_{sd}$. The blue part in the middle corresponds to supercurrent. (b) Corresponding plot of (a) for out-of-plane magnetic field $B_z$.

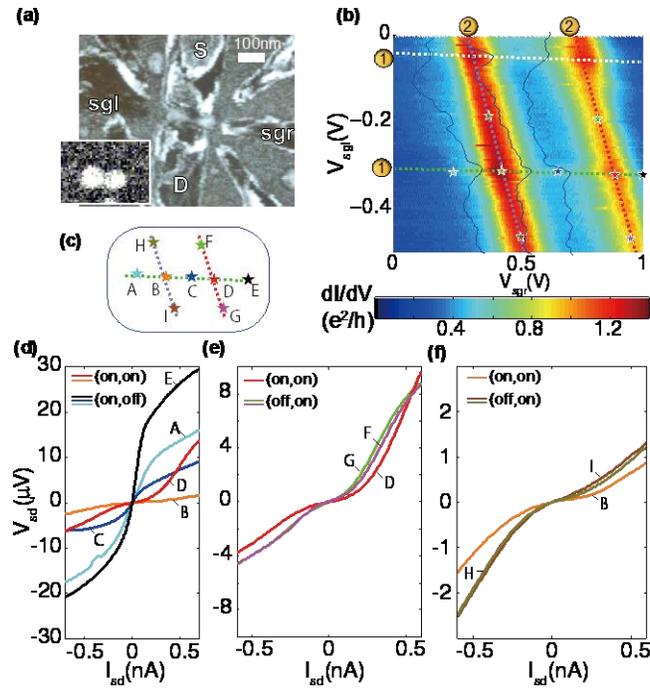

**Figure 4.** (a) A SEM micrograph of a double dot Josephson junction. The source/drain/sidegate electrodes are marked with S/D/sgr/sgl respectively. Note that only two of the electrodes surrounding the junction are used as sidegates. (b) Plot of zero-bias differential conductance as a function of two sidegates $V_{sgr}$ and $V_{sgl}$. Conductance peaks due to Coulomb resonance are highlighted by dashed lines, each accompanied by circles indicating the corresponding QD. 3 conductance traces taken along lines parallel to QD2 peaks are shown to highlight QD1 peaks. (c) Schematic illustration of the peak and mark positions in (b). (d)-(f) Examples of *V-I* traces taken at gate positions marked by stars along the green/red/blue dashed lines in (b), respectively.

**Supplemental Material : Superconducting Transport in single and parallel double InAs Quantum Dot Josephson Junctions with Nb-based Superconducting Electrodes**

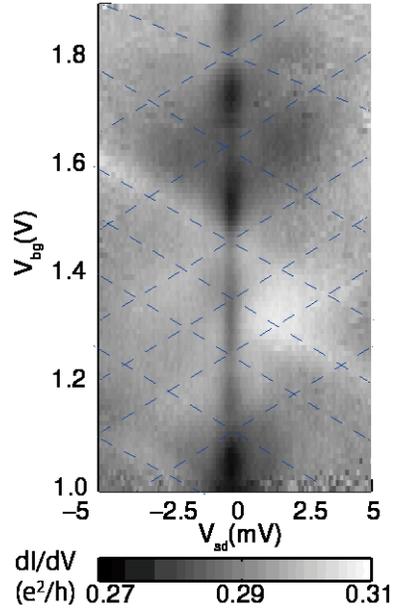

**Figure S1.** (a) Plot of differential conductance $dI_{sd}/dV_{sd}$ as a function of backgate and source-drain voltages. The data are from the same sample as shown in Fig. 1 in the main text. The differential conductance is recorded at the base temperature of ~300mK and external magnetic field $B_z$ =4T. Highlighted with dashed lines are Coulomb diamonds.

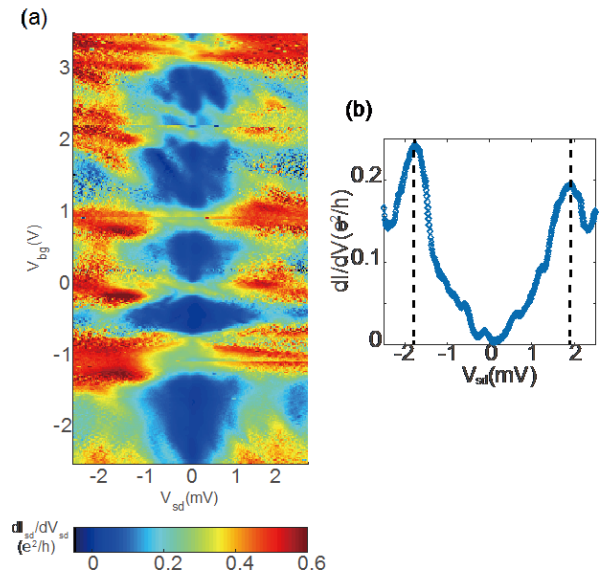

**Figure S2.** (a) Plot of differential conductance $dI_{sd}/dV_{sd}$ as a function of backgate and source-drain voltages. The data are from the same sample as shown in Fig. 4 in the main text. Dashed lines show the approximate positions of the peaks due to quasi-particle tunneling, from which $\Delta$~0.95meV is obtained. Note that resonance peaks of QD1 are not clearly observed here due to their low amplitude. Since the diameters of QDs are comparable, we estimate $U_1$~$U_2$=1 to 2meV. (b) A typical $V_{sd}$ dependence of $dI_{sd}/dV_{sd}$ extracted from (a) at $V_{bg}$=-0.5V.

|  | Al | Nb | NbTiN |
|---|---|---|---|
| $B_c$(T) | ~0.11 (at 40mK) | ~4.5 (at 300mK) | - |
| $\Delta$(meV) | ~0.13 | ~0.55 | ~0.95 |
| $T_{clead}$(K) | - | ~7.0 | - |

**Table S1.** Typical characteristics of the junctions with each superconducting material. $\Delta$ is derived from the separation of peaks due to quasi-particle tunneling. $B_c$ here is the applied perpendicular field at which this gap closes. Note that $T_{clead}$ is the critical temperature of the 200nm width lead, not of the junctions with QDs.